# Flow and Heat Transfer in Micro Pin Fin Heat Sinks with Nano-Encapsulated Phase Change Materials

Bahram Rajabifar

Hamid Reza Seyf

Yuwen Zhang[1]
Fellow ASME
zhangyu@missouri.edu

Sanjeev K. Khanna
Fellow ASME

Department of Mechanical and Aerospace Engineering

University of Missouri

Columbia, MO 65211

**Abstract**

In this paper, a 3D conjugated heat transfer model for Nano-Encapsulated Phase Change Materials (NEPCMs) cooled Micro Pin Fin Heat Sink (MPFHS) is presented. The governing equations of flow and heat transfer are solved using a finite volume method based on collocated grid and the results are validated with the available data reported in the literature. The effect of nanoparticles volume fraction (C = 0.1, 0.2, 0.3), inlet velocity ($V_{in}$ = 0.015, 0.030, 0.045 m/s), and bottom wall temperature ($T_{wall}$ = 299.15, 303.15, 315.15, 350.15 K) are studied on Nusselt and Euler numbers as well as temperature contours in the system. The results indicate that significant heat transfer enhancement is achieved when using NEPCM slurry as an advanced coolant. The maximum Nusselt number when NEPCM slurry (C = 0.3) with $V_{in}$ = 0.015, 0.030, 0.045 (m/s) is employed, are 2.27, 1.81, 1.56 times higher than the ones with base fluid, respectively. However, with increasing bottom wall temperature, the Nusselt number first increases then decreases. The former is due to higher heat transfer capability of coolant at temperatures over the melting range of PCM particles due to partial melting of nanoparticles in this range. While, the latter phenomena is due to the lower capability of NEPCM particles and consequently coolant in absorbing heat at coolant temperatures higher than the temperature correspond to fully melted NEPCM. It was observed that NEPCM slurry has a drastic effect on Euler number, and with increasing volume fraction and decreasing inlet velocity, the Euler number increases accordingly.

[1] Corresponding Author.

## Introduction

With rapid advancement in microfabrication techniques, fabrication of microscale devices for different applications, including cooling of electronic devices, has been become possible. Due to their high surface area per unit volume, compact size, and high heat transfer effectiveness, Micro-pin-fin heat sinks are an important class of heat transfer devices used in electronic cooling components. It consists of an array of fins, which extended from a base area and are closely constrained by the opposing wall and other sides of channel. In order to evaluate the performance of micro-pin-fin heat sinks, understanding the physics of flow and heat transfer phenomena in these systems are essential. In the recent years, several studies have been performed by previous researchers in order to determine hydrodynamics and thermal characteristics of micro-pin-fin arrays. Go et al. used flow-induced vibration of a micro-fin array to experimentally investigate the enhancement of heat transfer of the system [1]. Peles et al. [2] investigated a bank of micro pin fins in terms of its pressure drop and heat transfer characteristics and tried to optimize the geometrical and thermo-hydraulic parameters to reduce the overall thermal resistance of the system. They showed that thermal performance of microscale pin fin heat sinks provide a very low thermal performance and sparse and dense pin fin configurations may be employed to reduce the overall thermal resistance of the system.

Kosar et al. [3] obtained friction factor over intermediate size pin fin heat sinks experimentally. They reported deviations from long tube correlations at low Reynolds numbers and low fin height-to–diameter ratios. Kosar [4] evaluated the performance of five assorted MEMS- based pin fin heat sinks with different shapes, spacing and arrangements. It was found that the effects of arrangement, shape of pin fins, and spacing on heat transfer and friction factor were in agreement with the results of existing large-scale reported in the literature. Kosar and Peles [5] studied boiling inception, pressure drop and heat transfer of R-123 over a bank of micro pin fins at Reynolds numbers from 134 to 314. They examined that effect of end walls on heat transfer at different Reynolds numbers and realized that the end walls effect is negligible at Reynolds numbers greater than 100 and conventional scale correlations are reliable in this range but at low Reynolds numbers, the correlations underpredict the experimental data, significantly.

The single-phase pressure drop in an array of staggered micro-pin-fins was studied by Qu and Siu-Ho [6]. They suggested a new friction factor correlation, which showed good agreement with both diabatic and adiabatic data. Seyf and Feizbakhshi [7] using a three-dimensional conjugated heat transfer model investigated the effects of CuO and $Al_2O_3$ nanofluids on thermal performance of micro pin fin heat sinks with circular pins. They showed that Nusselt number increases with increasing volume fraction but the effect of particle size on thermal and hydrodynamic performance of heat sink is different for different nanofluids. The pressure drop and heat transfer in a copper micro square pin fin heat sink was experimentally investigated by Liu et al. [8] and showed that the increase of Reynolds number results into the increase of the pressure drop and Nusselt number of the system.

Enhancing the heat capability of coolant using NEPCMs is an innovative method for increasing the thermal performance of a heat transfer device such as heat sinks. In this method, the NEPCM particles are suspended in a coolant to construct phase-change slurry. The heat storage capability of coolant increases due to the phase change of PCM nanoparticles in the base fluid; hence it boosts the ability of coolant to absorb high heat fluxes. In general, NEPCM particles are made of a paraffin wax PCM core which is covered by a cross-linked polymer shell. The shell is designed to be flexible enough in order to tolerate volume changes that expected to accompany phase transitions [9]. Recently, utilization of micro / nano encapsulated phase material (MEPCM / NEPCM) slurries to enhance heat transfer rates of cooling devices attracted a great attention [9-15]. Sabbah et al. [9] conducted a numerical study to understand the effect of the outlet and inlet temperatures and melting temperature range of the PCM on the performance of a microchannel heat sinks with water-based slurry. Kuravi et al. [12] studied the effects of nanoencapsulated PCM slurry in manifold microchannel heat sinks. They showed that the slurry coolant causes higher thermal performance of heat sink compared to pure fluid. Hao and Tao [10] showed the improving effect of utilizing micro and nano-sized PCM particle in microchannel heat sinks. Wang et al. [11] experimentally compared the heat transfer coefficient of a laminar flow in a horizontal tube with and without MEPCM particles and reported significant heat transfer enhancement in case with NEPCM slurry coolant.

Sabbah et al. [13] using numerical simulation showed considerable enhancement in natural convection heat transfer when MEPCM was employed inside a rectangular cavity (up to 80%). Kondle et al. [14] studied the laminar flow heat transfer in microchannels with rectangular and circular channels to realize the effect of using PCM particles and showed a significant cooling performance enhancement could be obtained by utilizing PCM particles. Seyf et al. [15] studied the hydrodynamics and thermal performance of a NEPCM slurry-cooled tangential microchannel heat sink. They showed that mixture of octadecane as nanoparticles and poly-a-olefin (PAO) as the base fluid could cause lower thermal resistance compared with pure poly-a-olefin. They also studied the flow and heat transfer over a square cylinder and showed that increasing the volume fraction enhances both local heat transfer and shear stress over the surface of the cylinder [16].

Investigations into the use of NEPCM slurry in cooling devices are still embryonic and much more study is required in order to better understand the effect of these types of coolants on the performance of these devices. This paper is presenting a conjugated heat transfer model to study the effect of NEPCM utilization on thermal and hydrodynamic performance of MPFHS. The effect of volume fraction, inlet velocity as well as bottom wall temperature as heat source on flow and heat transfer characteristic in the system are studied in detail. The thermophysical properties of NEPCM slurry especially specific heat are strong functions of temperature and thus the behavior of the thermal performance of device varies with temperature. We investigated the thermal behavior of the device for different heating conditions and found that there is an optimum heat transfer, which is function of inlet velocity, volume fraction and the applied heat to the system.

**Problem Definition and Method of Solution**

A conjugated heat transfer model has been developed to study the flow and heat transfer in a MPFHS with rectangular pins (see Fig. 1) and PCM slurry as coolant. For simplicity, only one symmetrical part of heat sink consisting micro pin fins is adopted in our simulation. The side ($W_{fin}$), height (H), and gap between ($W_c$) silicon pin fins are 559 μm, 3 mm, and 241 μm, respectively. The height of computational domain is similar to height of pins and it contains three rectangular volumes, i.e., the central region which contains the pins and heated wall, the flow developing inlet block and the outlet block with lengths of nearly 35, 9, and 26 times of pin side, respectively. The length of inlet and outlet blocks are cautiously defined based on the used values in literature [17] and previous studies done by the authors [15, 18] to make sure that the flow becomes fully developed before entering the central computational region and to avoid the influence of back streams, respectively. The staggered arrangement is in form of an equilateral triangle with longitudinal and transverse pitch-to-side ratios of $S_L$ = 0.5657 and $S_T$ = 0.5657, respectively. The used dimensions in this study are defined based on the ones used by Liu et al. [8] to facilitate validation of the results. The governing equations of conservation of mass, momentum and energy for an incompressible flow can be written as

$$\frac{\partial(u\psi)}{\partial x}+\frac{\partial(v\psi)}{\partial y}+\frac{\partial(w\psi)}{\partial z}=\frac{\partial}{\partial x}\left(\Gamma_\psi \frac{\partial\psi}{\partial x}\right)+\frac{\partial}{\partial y}\left(\Gamma_\psi \frac{\partial\psi}{\partial y}\right)+\frac{\partial}{\partial z}\left(\Gamma_\psi \frac{\partial\psi}{\partial z}\right)+S_\psi \tag{1}$$

where $\psi$ stands for the velocity components and temperature, i.e., $u, v, w$ and $T$. $S_\psi$ and $\Gamma_\psi$ are corresponding source and diffusion terms, respectively which are defined in reference [7]. The continuity equation can be obtained when $\psi = 1,\ \Gamma_\psi = 0,\ \mathrm{S}_\psi = 0$.

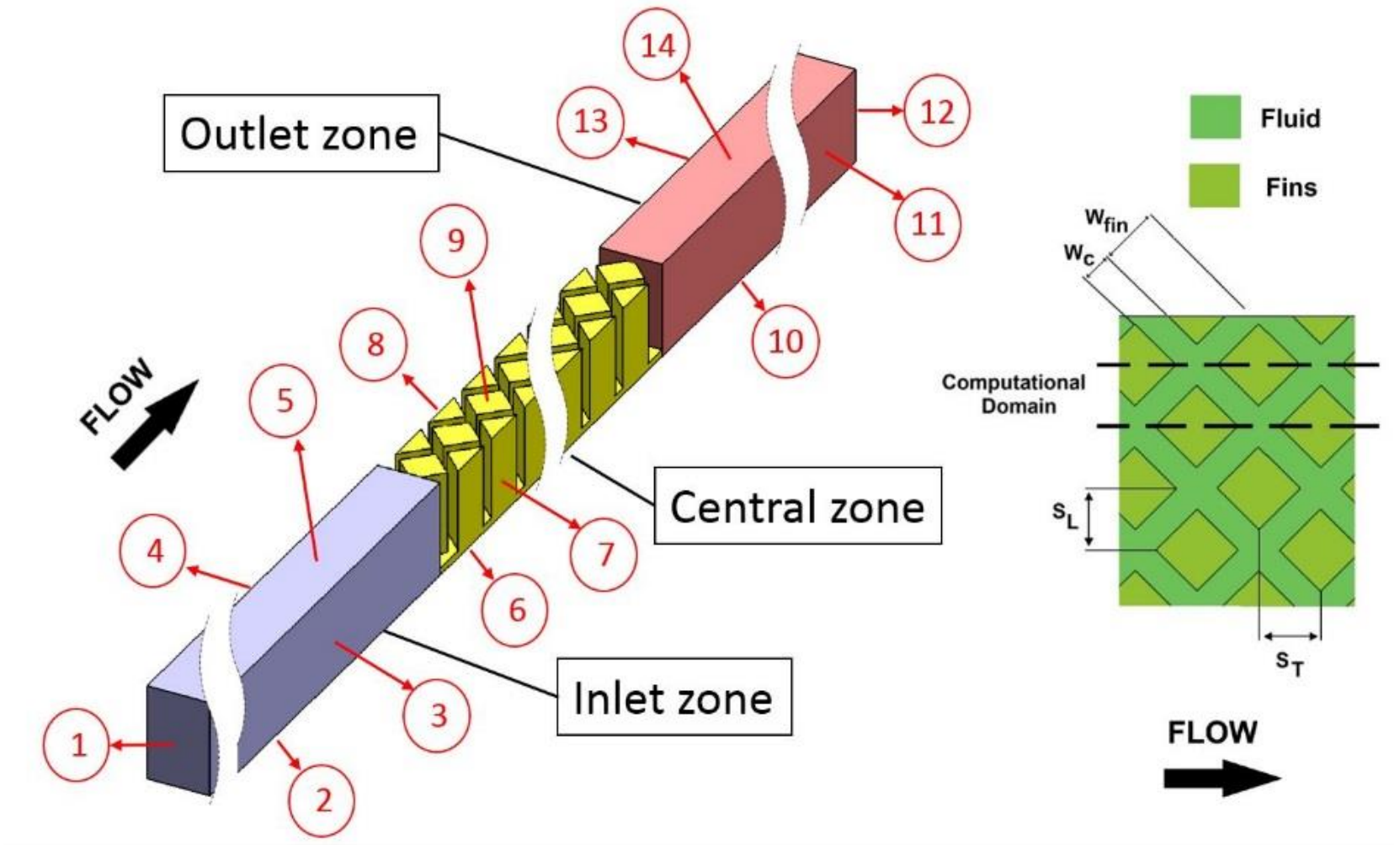

Figure 1. Schematic of the computational domain

A combination of outlet, wall, inlet, and symmetry boundary conditions are applied in the computational domain. Each of the surfaces are denoted by a number in the schematic shown in Figure 1. Constant temperature boundary condition with different values ranging from 296.15K to 350.15 K is applied on the bottom surface of the central region (surface 6) whereas the top wall of domain is insulated (surfaces 5, 9, and 14). At the inlet (surface 1), a constant temperature (296.15 K) and a uniform velocity profile are considered. The rectangular micro pin fins and channels are treated as no-slip boundary condition and at the outlet (surface 12) the static pressure is fixed and the remaining flow variables are extrapolated from interior of computational domain. Lateral surfaces of the computational domain (surfaces 3, 4, 7, 8, 11, and 13) are treated as symmetric boundaries and the remaining ones (surfaces 2, and 10) are considered adiabatic. For more details about the equations used for each boundary please refer to [7].

The base fluid of the utilized slurry is water and the added NEPCM particles are made of n-octadecane phase change material and their average diameter, melting point, density, specific heat, thermal conductivity, melting range, and latent heat are considered to be constant and equal to 100 nm, 296.15 K, 815 kg/m$^3$, 2000 J/kg.K, 0.18 W/m.K, 10 K and 244,000 J/K, respectively [12, 15]. The volumetric concentrations are used in this study include C= 0.1, 0.2, and 0.3 which are intended to be below 0.3 to provide Newtonian fluid characteristics [19]. By using and comparing different profiles to represent specific heat of PCM nanoparticles in order to investigate their effect on the results accuracy, Alisetti and Roy [20] proved that there is no significant preference and the results would not vary more than 4%. Therefore, the specific heat of NEPCM particles is represented by a sine profile as shown in Figure 2. The specific heat of the particles for temperatures in melting range ($T_{mr} = T_2 - T_1$), first increases to a maximum value and then decreases. Specific heat of the nanoparticles in temperatures above $T_2$ ($c_{p,s}$) and below $T_1$ ($c_{p,l}$) are assumed constant and equal. The inlet Reynolds number of slurry flow in MPFHS is intended to be less than 100 to fulfills the presence of laminar flow in the channels [7]. The flow is incompressible and depletion layer effect is not considered [21, 22]. NEPCM particles distributed homogeneously all over the slurry volume and they are moving at the same speed of the local flow. Viscous dissipation and radiation terms are also neglected. In addition, the micro-convection caused by the particle–particle, particle–wall, and particle–fluid interactions is assumed to be negligible. The effect of shell material of spherical shape encapsulated particles are assumed to be neglected and the particles melt instantaneously without temperature gradient formation inside the particle [12]. This paper is the continuation of our previous studies in this area and the readers are referred to our previous publications for details about assumption and relations for temperature dependent effective thermophysical properties of slurry [15, 18, 23].

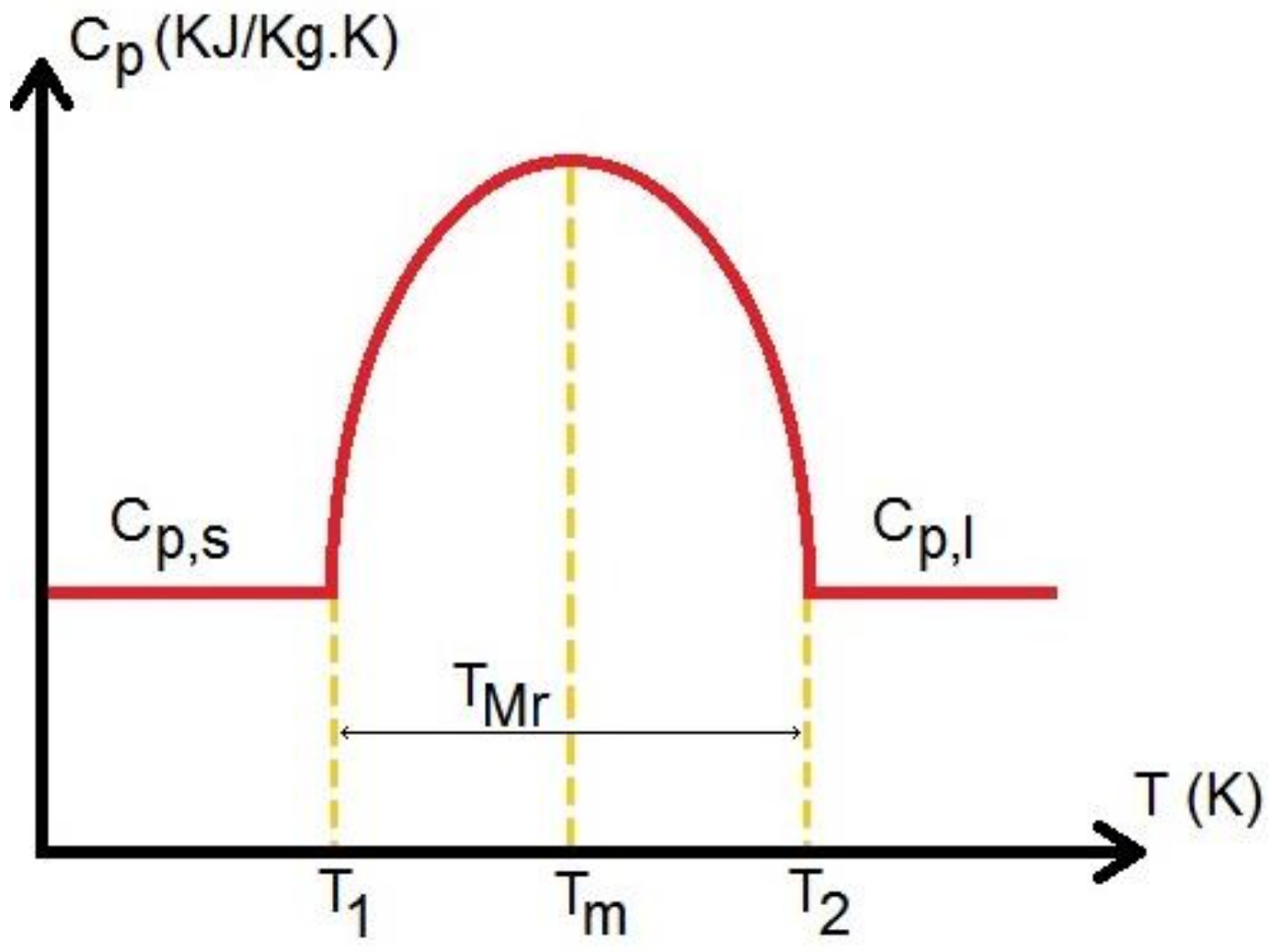

Figure 2. The specific heat of NEPCM particles is a function of temperature and is represented by a sine profile.

A validated code [7, 24] based on finite volume method and SIMPLE algorithm [25] has been employed as the numerical solver. Second order and QUICK schemes [24, 26] were used for discretization of diffusive and convective terms, respectively. The Fluent commercial package was employed together with user defined functions written in C++ to implement the property of NEPCM into the problem. In order to avoid velocity-pressure decoupling problem, the velocity components in discretized continuity equation are calculated using Rhie-Chow [27] interpolation technique. More details about numerical method can be found in [7, 24]

The maximum flow velocity in fin channel can be estimated using:

$$u_{\max} = \frac{Q_f}{A_{\min}} \tag{2}$$

where, $A_{min}$ is the minimum transverse flow area of square micro pin fin, which can be calculated from:

$$A_{\min} = WH \frac{W_c / \sin 45^o}{2S_T} \tag{3}$$

The Reynolds number based on the minimum width is given by:

$$\mathrm{Re}_c = \frac{\rho_{slurry} u_{\max} D_c}{\mu_{slurry}} \tag{4}$$

where subscript "slurry" indicate the effective properties of NEPCM slurry. The minimum channel width is determined by:

$$D_c = \frac{2HW_c \,/\sin 45^o}{H + W_c \,/\sin 45^o} \tag{5}$$

Euler number [7] presents the dimensionless pressure drop as:

$$Eu = \frac{2\Delta p}{\rho_{fm} U_m^{\ 2} N} \tag{6}$$

where $\rho_{fm}$ is mean coolant density, N is number of pin row, Δp is the pressure drop in the central region containing pin fins, and $u_m$ is mean velocity in the minimum cross section.

The heat flux of the hot wall can be calculated as follows:

$$q = \frac{(\rho c_p)_{nf} .u_{in} .A_{in} .(T_{out} - T_{in})}{A_h} \tag{7}$$

where, $A_{in}$ and $A_h$ denote the area of inlet and the base area of hot wall, respectively. $u_{in}$ is the inlet velocity and $T_{out}$ and $T_{in}$ are outlet and inlet bulk fluid temperatures, respectively.

The overall Nusselt number of the MPFHS is determined as follows

$$Nu = \frac{qD_h}{\left(T_h - \frac{(T_{in} + T_{out})}{2}\right) k_f} \tag{8}$$

where, $T_h$, $D_h$, and $k_f$ are temperature of bottom wall, the hydraulic diameter of the pin cross-section and fluid thermal conductivity, respectively.

**Grid independency and Validation**

An unstructured grid of tetrahedral volume elements were used for the central region that contain pins, while two structure grids were used for inlet and outlet, and blocks. Equiangular skewness of the tetrahedral grids are checked to ensure that the used grids are high quality. Four grids with different size of 836,125 (coarse), 1,298,211 (intermediate), 2,368,659 (fine) and 3,172,277 (very fine) are used to study the independency of the solution to number of grids. The mesh of the computational domain is uniformly refined between considered grid arrangements. Convergence is defined based on the temperature, velocity components, and pressure absolute errors of $10^{-8}$, $10^{-7}$, and $10^{-5}$, respectively. No pre-conditions are considered to reduce the iteration numbers. Table 1 presents the values and percentage difference of Nusselt and Euler numbers for the studied grid sizes for bottom wall temperature of 299.15 K. As seen the maximum difference between Nusselt and Euler numbers obtained based on fine and very fine grids are 2.01% and 2.28%, respectively hence in this paper, the fine grid is selected to conduct the simulation.

Table 1. Girds independency study

| $T_{wall}$ (K) | Number of grid | Euler Number | diff (%) | Nu | diff (%) |
|---|---|---|---|---|---|
| 299.15 | 836,125 | 2.118 | - | 5.072 | - |
| 299.15 | 1,298,211 | 1.793 | 15.37% | 4.189 | 17.40% |
| 299.15 | 2,368,659 | 1.593 | 11.14% | 3.710 | 11.42% |
| 299.15 | 3,172,277 | 1.561 | 2.01% | 3.626 | 2.28% |

It is worth mentioning that the code used in this paper has been validated against several experimental data [7, 24]. However, we further validate the code and present results by comparing the reported experimental data by Liu et al [8] with numerical results of the code. The experimental device consist of a copper micro pin fin heat sink cooled by deionized water with transfer area of $20 \times 20$ mm$^2$ with 3 mm height and $559 \times 559$ μm$^2$ cross section. Constant heat flux was applied to the bottom of the heat sink using eight 300W power cartridge heaters inserted from the bottom of the heat source made from pure copper while the top surface of system was insulated with Aspen Aerogel insulation for minimizing heat loss from top surface. The range of Reynolds numbers studied in the experiment was from 60 to 800 so the flow inside of the system was laminar, transition and turbulent flow based on Reynolds number and average Nusselt number. Liu et al. [8] obtained the value of 300 for transition Reynolds number based on the calculated average Nu variation in different Reynolds numbers.

We have simulated only one symmetrical part of MPFHS used in the experimental work of Liu et al. [8] consisting of pins and surrounding fluid defined as solid and fluid zones, respectively. Constant temperature and velocity boundary conditions at inlet and constant temperature at the bottom wall of the heat sink similar to the values used in the experiment [8] were used. The surface of pin fins were treated as no slip boundary condition and at the outlet the constant static pressure equal to atmospheric pressure was used; the remaining variables were extrapolated from the interior of computational domain. Validation of laminar flow was performed at different heat fluxes and Reynolds numbers. Table 2 shows the comparison between experimental and numerical results for laminar flow. The agreement between numerical results and experimental data is very satisfactory with less than 3.2% difference. Considering possible deviation of the manufactured micro-pin-fin dimensions from the nominal values used in numerical model as well as experimental uncertainties (dimensions: 0.5%, Temperature: 2%, Heat flux: 6.3%), the obtained numerical results well agree with experimental data. Furthermore, as the second validation we validate our results with a set of experimental data [28] and another set of numerical simulations results [12] for flow of PCM slurry in a tube which is presented in Figure 3 and illustrates a proper agreement.

Table 2. Comparison between experimental and numerical result

| $V_{in}$ (L/hr) | Heat Flux (W/m2) | $T_w$ - Numerical (K) | $T_w$ - Experimental (K) | %error |
|---|---|---|---|---|
| 5.693 | 170,110 | 309.18 | 307.50 | 0.55% |
| 5.693 | 521,000 | 339.46 | 334.09 | 1.61% |
| 5.693 | 664,300 | 351.42 | 345.28 | 1.78% |
| 19.075 | 208,748 | 303.19 | 298.92 | 1.43% |
| 19.075 | 368,720 | 310.67 | 304.27 | 2.10% |
| 19.075 | 639,800 | 323.18 | 313.38 | 3.13% |

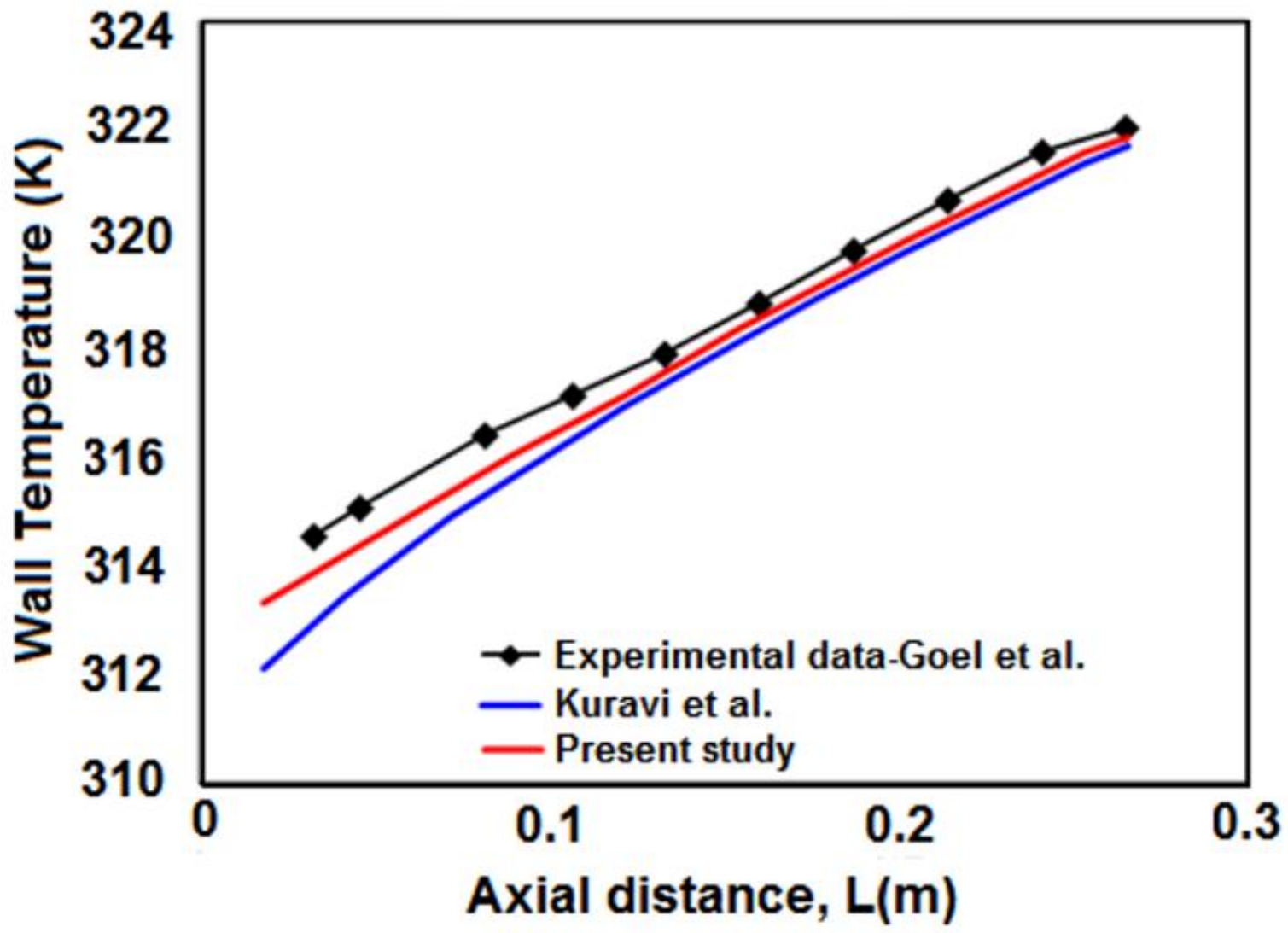

Figure 3. Comparison of result using current model, numerical model in [12] and experimental data in [28]

**Results and Discussions**

The effects of different parameters, such as inlet velocity ranging from 0.015 to 0.045 m/s, volume fraction of NEPCM particles ranging from 0 to 0.3, and bottom wall temperature ranging from 299.15 to 350.15 K, on temperature distribution, Euler and Nusselt numbers are studied. The effects of inlet velocity on temperature distribution in both fluid and solid regions of computational domain are shown in Figure 4 for C= 0.3 and $T_w$=315.15 K. The applied heat at the bottom of the heat sink spreads to the fins so that the temperature is high at the bottom of the heat sink but it decreases along the height of micro pin fins because of the interaction of coolant with pin fins that cools the pin fins. It can be seen that with increasing inlet velocity the temperatures of coolant and solid region of heat sink decreases because of lower thermal boundary layer thickness on the fins and bottom surface and consequently higher heat transfer rates from fins to the coolant. It is worth noting that with increasing inlet velocity, the average velocity of coolant in the heat sink increases and therefore, the heat transfer between solid surfaces and working fluid increases because of generation of thinner thermal boundary layers on the pin fins. The convection heat transfer in the coolant is comprised of two different mechanisms, i.e., energy transfer induced by the bulk motion of working fluid, and energy transfer due to interior thermal diffusion in the coolant. With increasing inlet velocity, the mean coolant velocity in the system increases and the forced convection will dominate the heat transfer in the system and therefore, the coolant may transfer more heat without considerable temperature increase. On the other hand, when inlet velocity is relatively low, the coolant has more chance to absorb and spread heat and diffusive heat transfer would play a more significant role in the heat transfer. It can be also seen that for V=0.015 m/s the temperature field in both coolant and fins become fully developed after the twentieth fin while for other cases the temperature in the system does not reach to fully developed condition. Similar behaviors have been observed for other bottom wall temperatures and volume fraction of NEPCM particles.

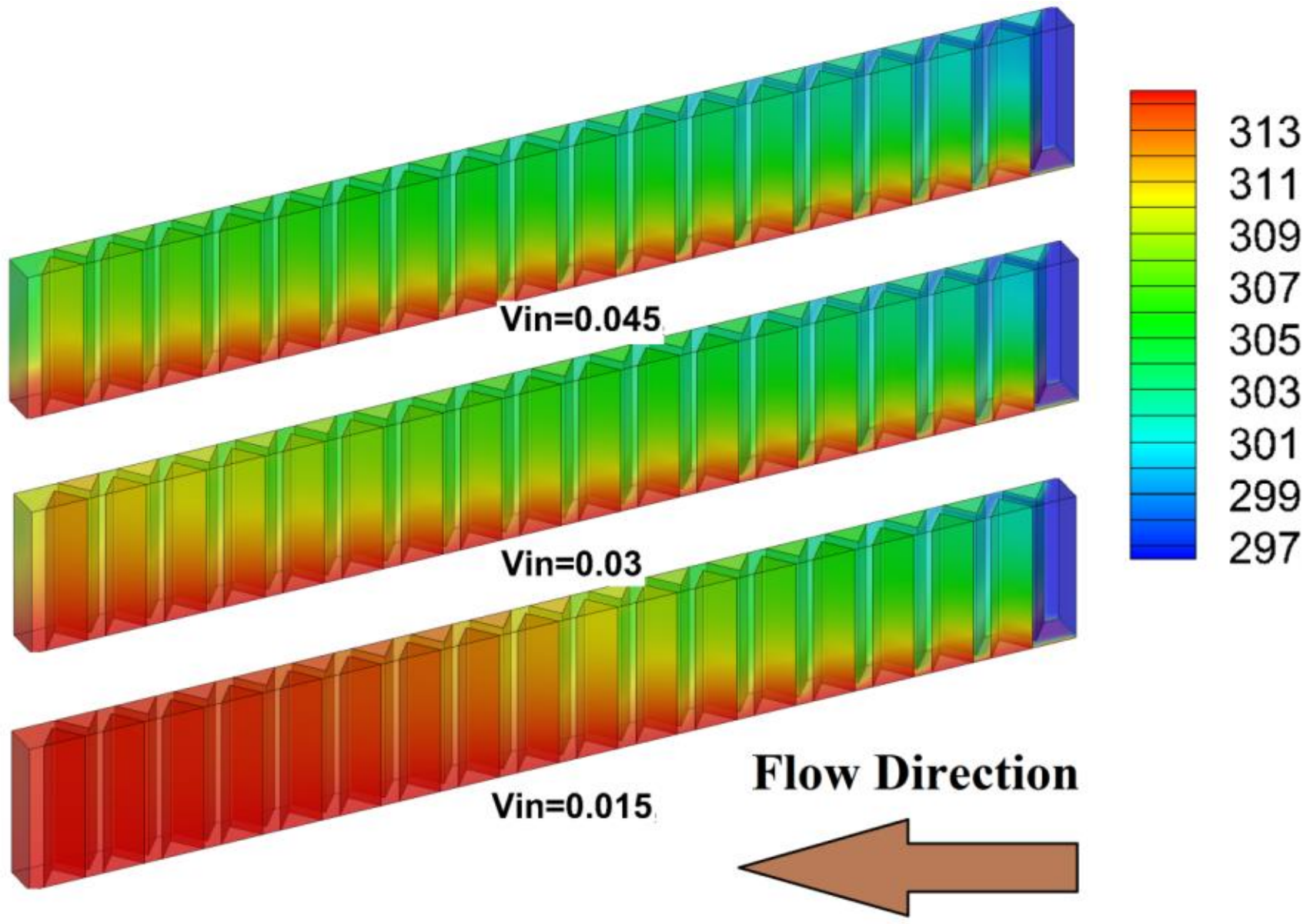

Figure 4. The temperature distribution in the system at three different inlet velocities for C= 0.3 and $T_w$=315.15 K. (Unit: K)

Figure 5 illustrates the effect of the volume fraction of NEPCM particles on temperature distribution in the computational domain at $T_w$ = 315.15 K and $V$ =0.045 m/s. As expected, due to higher heat capacity of the slurry coolant, using NEPCM slurry as coolant leads to lower fin temperature and consequently higher heat transfer coefficient; with increasing the volume fraction of nanoparticles, the reduction in temperature intensifies, which is an indication of higher heat transfer at higher volume fractions. Furthermore, as seen using NEPCM slurry cause the thermal boundary layer thicknesses on the fins and bottom surface decrease because the slurry NEPCM particles act as heat sink and slows down the thermal boundary layer thickening. It can also be seen that the temperature distribution in fluid region becomes smoother as volume fraction increases. This is due to the fact that NEPCM particles increase the viscosity of slurry, which makes the coolant more stable and consequently dampen the complex flow fields.

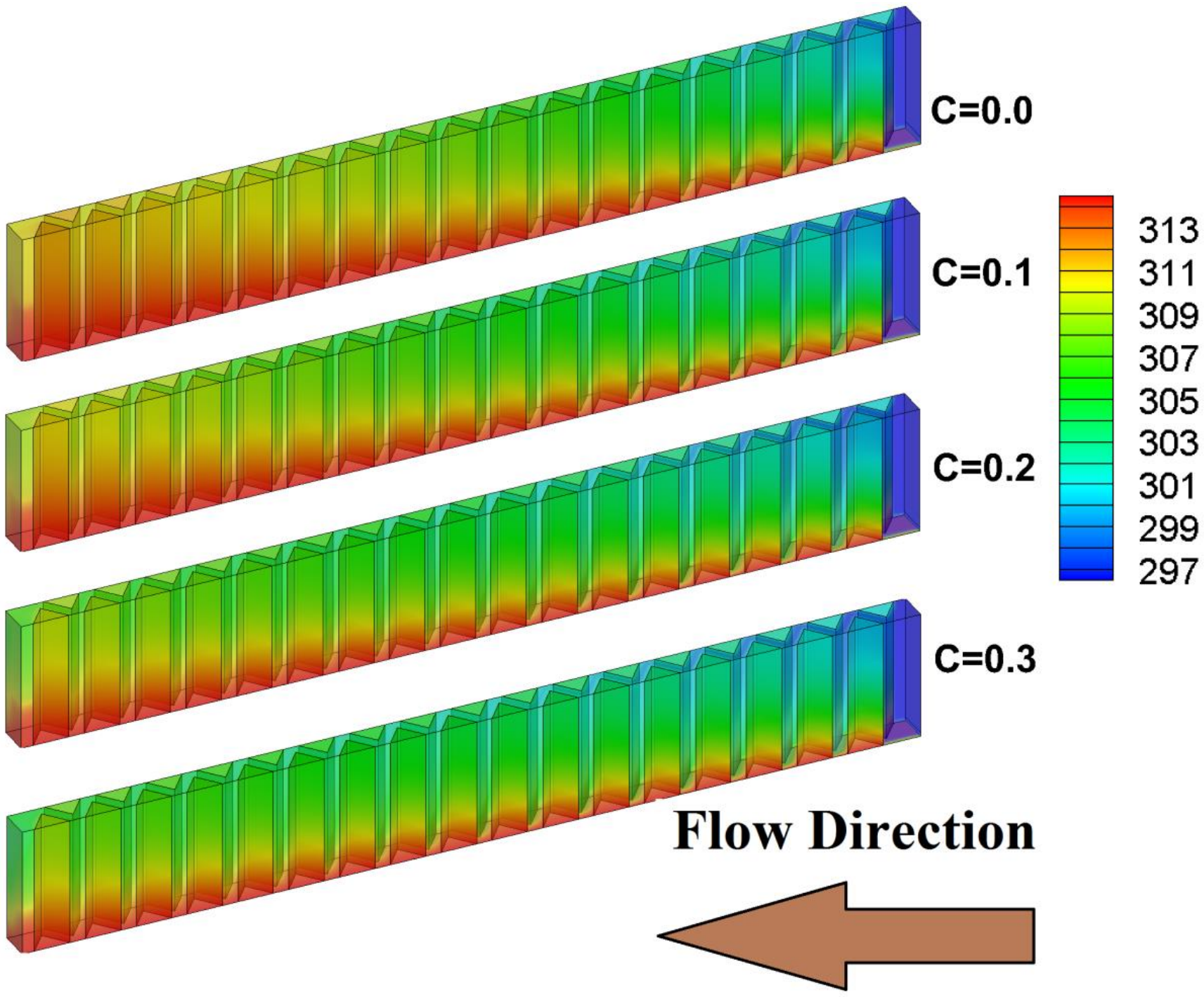

Figure 5. The temperature distribution in the system at three volume fraction for $V_{in}$ = 0.045 m/s and Tw=315.15 K. (Unit: K)

One of the main questions in designing a cooling device with NEPCM slurry is how much more performance can be gained by using NEPCM slurry as a coolant. To answer this question, the combined effects of volume fraction, bottom wall temperature, and inlet velocity on Nusselt number for different inlet velocities is shown in Figure 6 (a)-(c). As explained previously, increasing the volume fraction of NEPCM causes the thermal boundary layer on the solid surfaces to decrease and the effective heat capacity of coolant to increase. Therefore, the heat transfer and consequently Nusselt number increases as shown in Figure 6 (a)-(c). Moreover, with increasing the inlet velocity at a constant volume fraction of NEPCM, the heat transfer coefficient increases due to reduction in thickness of thermal boundary layer. Therefore, one can conclude that higher volume fractions and inlet velocities result in more effective cooling. It is worth noting that, when the inlet temperature is less than solidus temperature ($T_1$), three different heat transfer regions and relevant rates are observed along the fluid flow path. At the first region, where the NEPCM particles are mostly in solid state, the heat capacity of the coolant is comparable with the base fluid and the effective thermal conductivity is slightly lower than the one of the base fluid. Therefore, not a considerable change in cooling performance of the system in comparison with the original condition is expected. In this region, which is located near the inlet of the heat sink, the main heat transfer mechanism is based on the huge temperature difference between hot solid surfaces and cold coolant. At the second region, which is located after the first region, the NEPCM particles undergo the state transition (solid to liquid) and the latent heat contribution to the process, significantly enhances the effective heat capacity of the coolant. This region is the main reason behind the improved cooling performance of the systems with NEPCM slurries, in which the higher effective heat capacity of the coolant rather than conventional types of coolants, improves the system cooling performance, considerably. Finally, at the third region, the NEPCM particles are mainly expected to be in liquid state and the effective heat capacity of the coolant is similar to the first region. Therefore, due to higher bulk temperature of the coolant in this region, which results in smaller temperature gradient between the pin fin surfaces and working fluid, the minimum cooling performance of the system occurs in this region. This justifies the observed cooling capability of the coolant which increases and then decreases.

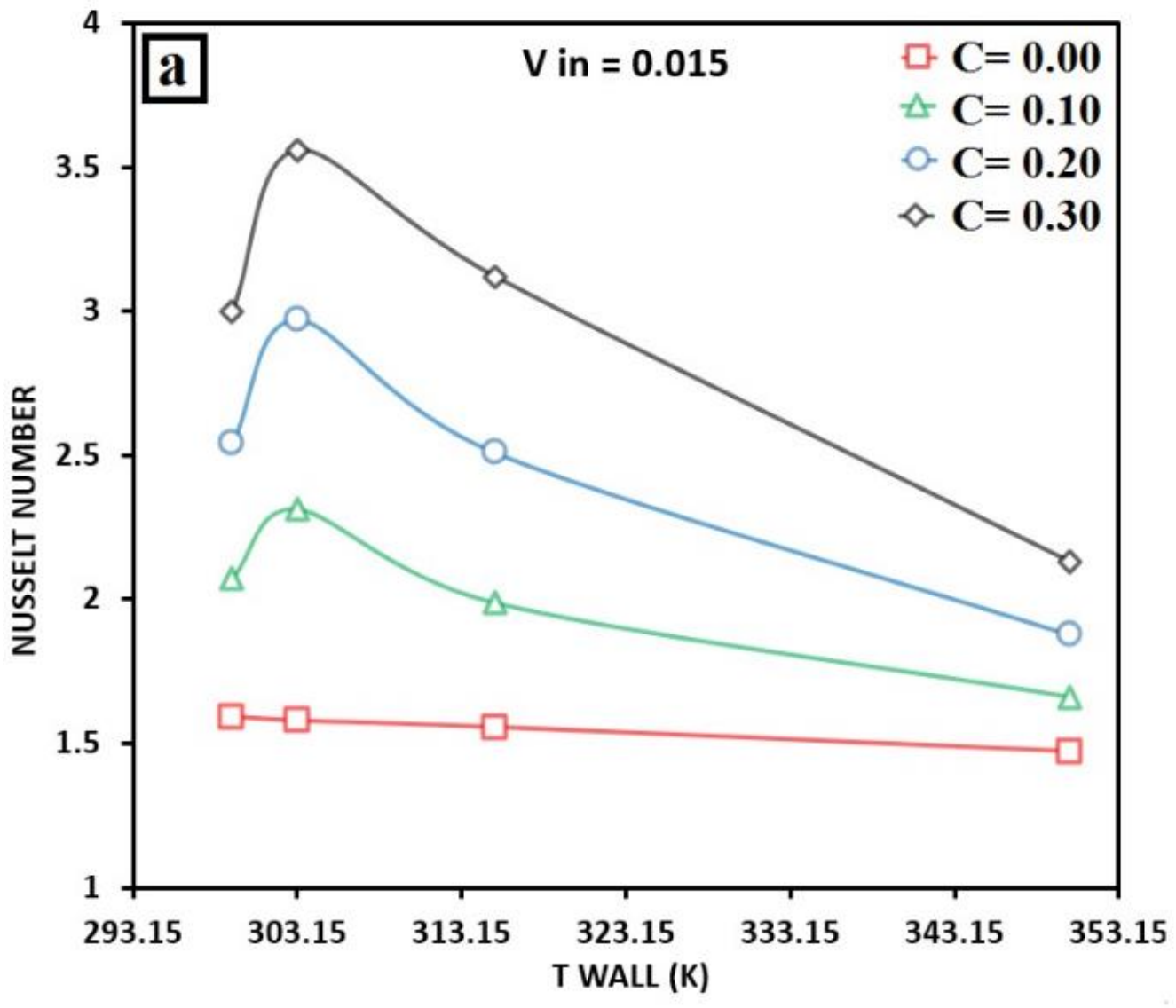

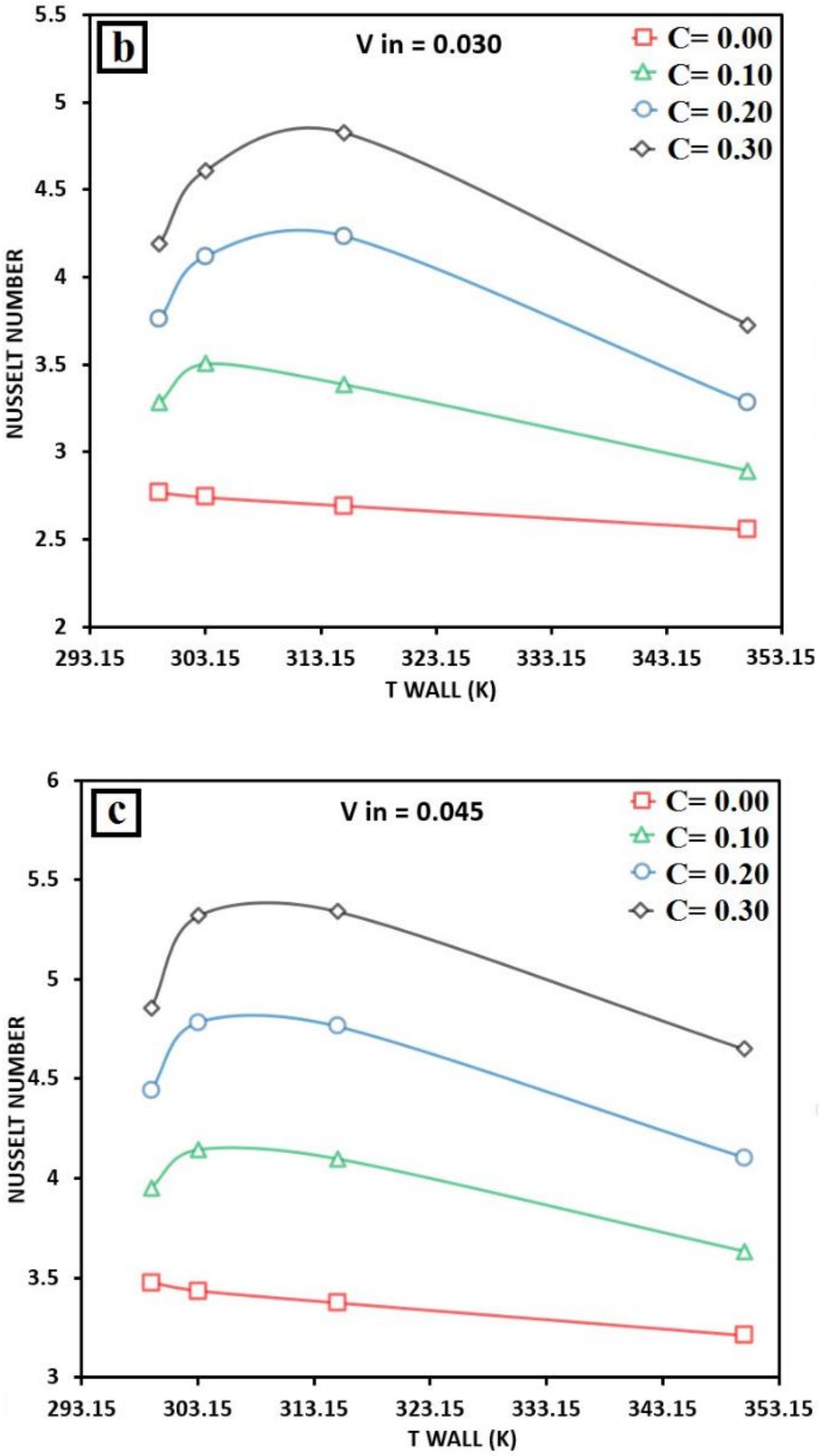

Figure 6. Nusselt number of the system (cooling performance) as a function of volume fraction of NEPCM particles and various inlet velocities.

As seen from Figure 6 (a)-(c), the bottom wall temperature has a significant effect on Nusselt number. For the case of pure water, at constant inlet velocity and temperature, with increasing wall temperature the Nusselt number decreases almost linearly. For the cases with NEPCM slurry as coolant, as wall temperature increases, the Nusselt number first increases and then decreases. This phenomenon is due to the growing and decaying trends of effective specific heat of slurry with respect to the temperature at different coolant temperature regions. At the inlet of heat sink, the temperature of slurry is constant ($T_1$) and PCM nanoparticles are solid so the specific heat is minimum. With increasing the wall temperature, the coolant temperature increases accordingly and depending on its value different trends are observed in specific heat of slurry and consequently Nusselt number. For instance, as seen in Figure 7, at constant inlet velocity of 0.015 m/s and nano-pcm volume fraction of 0.3, for the wall temperature of 299.15 K due to high local temperature of coolant, certain amount of nanoparticles are melted especially in areas near the bottom wall and the pin fin surfaces. Consequently, in this region local and volume weighted average of specific heat increases and it acts to increase the heat transfer and cooling performance of the system in two ways: first it increase the heat capacity of the coolant and this lets the coolant absorb more amount of heat. Then the higher heat capacity of the melting NEPCM resists the temperature increase trend and slows the temperature increase rate. These two reasons improve the cooling performance of the system and convert the pure fluid coolant to a coolant of high quality. However, with increasing wall temperature from 299.15 K to 303.15 K, the coolant temperature increases more which causes higher average specific heat of slurry due to larger amount of melted nanoparticles and consequently enhancement in Nusselt number as shown in Figure 6 (a)-(c). Increasing the wall temperature beyond 303.15 K causes a decaying trend for the Nusselt number because of very high temperature of slurry in most parts of the system, which causes full melting of NEPCM that results in a lower specific heat of slurry and consequently lower Nusselt number. Similar trends are observed in contour of specific heat for different volume fractions and inlet velocities. The effects of velocity on contour of specific heat of slurry for wall temperature of 315.15 K and volume fraction of 0.3 are shown in Figure 8. It can be seen that the velocities in these cases has a desirable effect on melting of nanoparticles and consequently enhancement of slurry specific heat hence and as shown in previous part, the Nusselt number increases with increasing inlet velocity in these cases (see Figure 6(a)-(c)).

Since, NEPCM particles at their melting range are ready to absorb a considerable latent heat, slurries with greater nanoparticles volume fraction apparently are able to absorb more energy than the slurries with smaller volume fractions and at the same time, as illustrated in Figure 5, at the same bottom wall temperature, the slurries with greater volume fraction enjoy a lower bulk temperature. In other word, a specified average bulk temperature of slurry in case of greater nanoparticles volume fraction slurries, is achieved at a higher bottom wall temperature in comparison with slurries with smaller nanoparticles volume fraction. Therefore, if the nanoparticles volume fraction is greater, the average bulk temperature in which majority of nanoparticles are in their melting range would occur in a higher bottom wall temperature and as a result, as illustrated in Figure 6, the maximum Nusselt number of the heat sinks with slurries with greater nanoparticles volume fraction shows a slight tendency to occur at higher bottom wall temperatures.

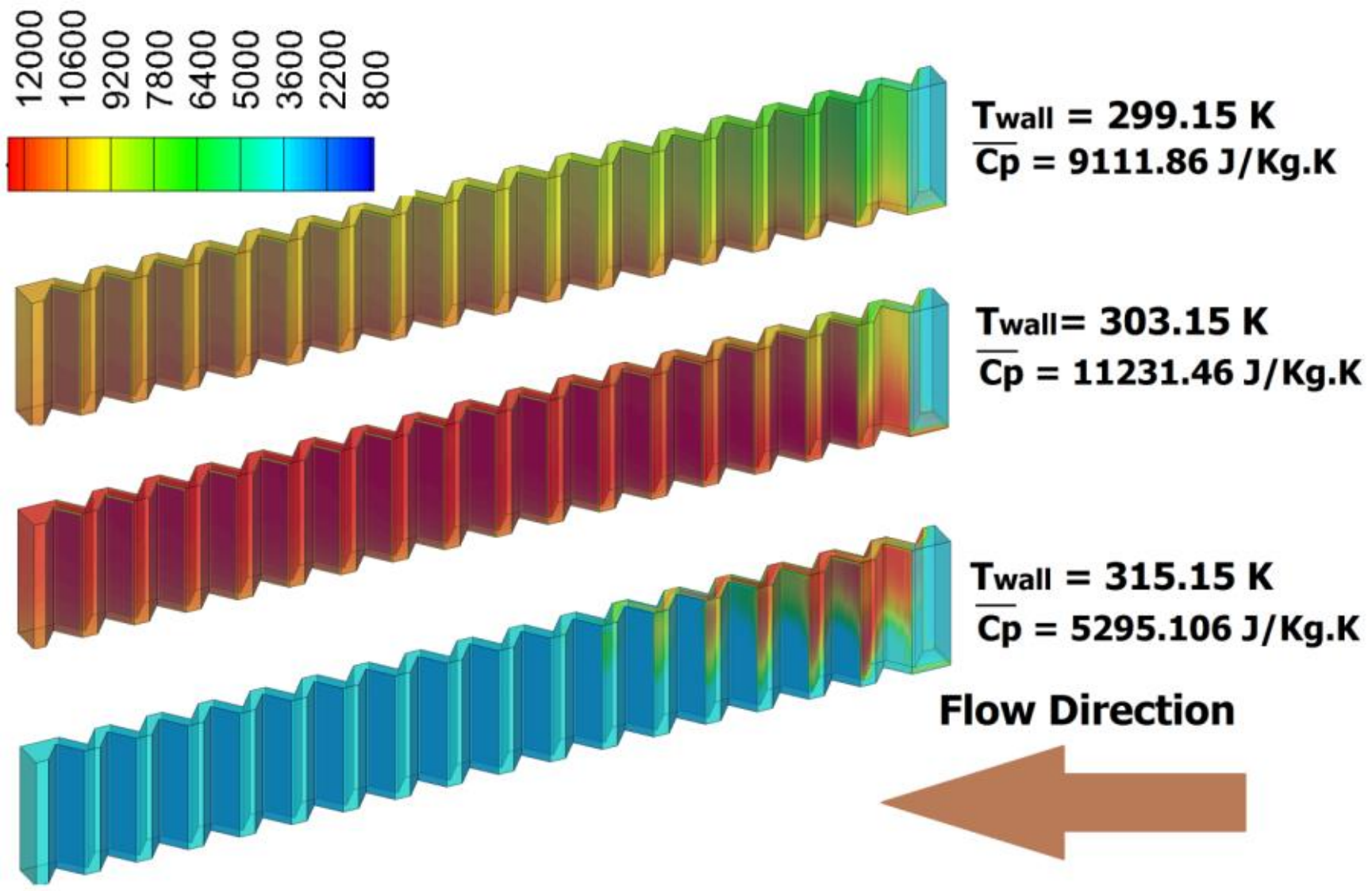

Figure 7. The distribution of specific heat of slurry in the system at constant inlet velocity of 0.015 m/s and volume fraction of 0.3 (Unit: J/kg.K)

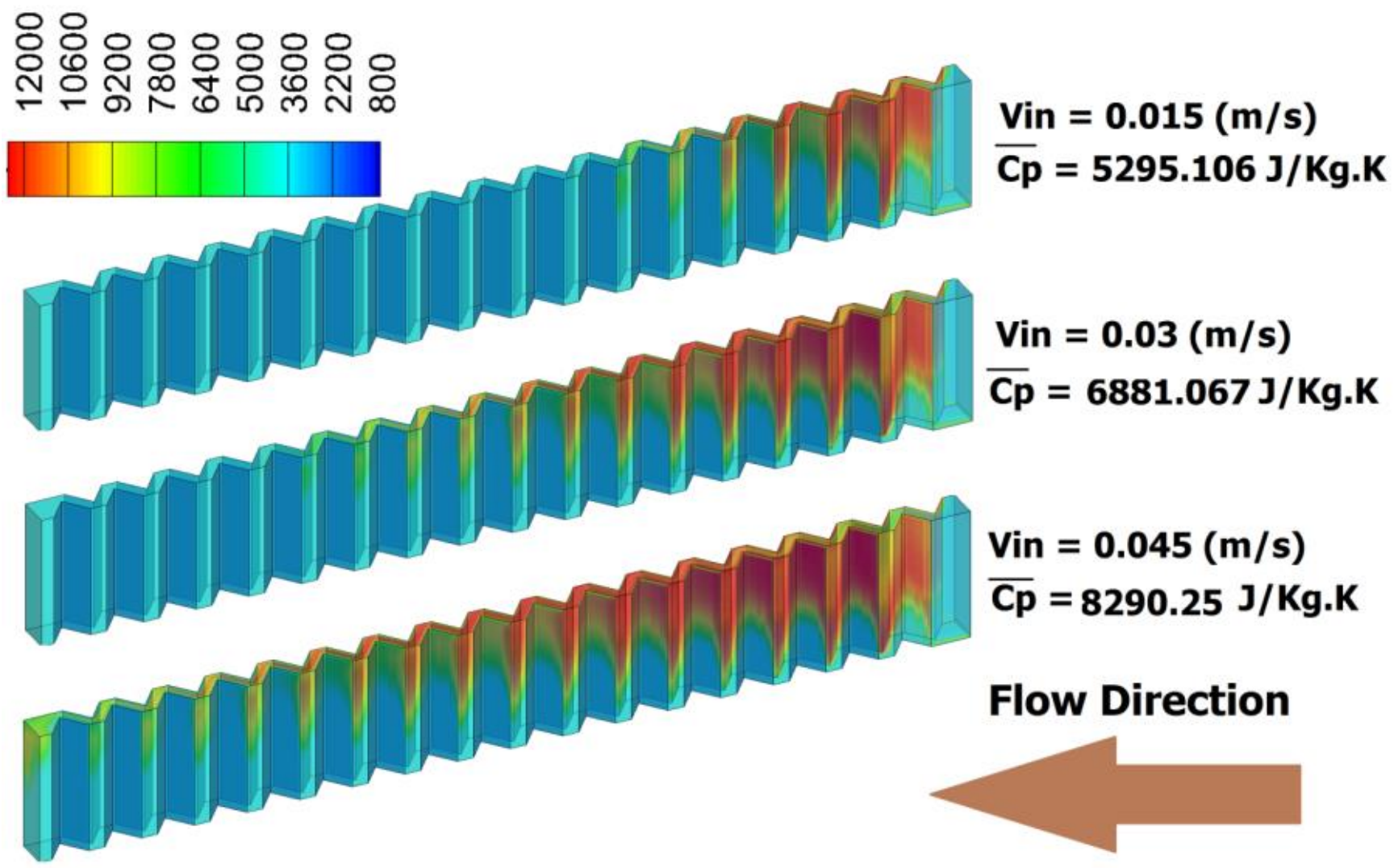

Figure 8. The distribution of specific heat of slurry in the system at bottom wall temperature of 315.15 K and volume fraction of 0.3 (Unit: J/kg.K)

Figure 9 illustrates the effects of volume fraction, inlet velocity and bottom wall temperature variations on Euler number. It can be seen that by decreasing the inlet velocity, the Euler number increases and the sensitivity of Euler number to volume fraction decreases at higher inlet velocities. Furthermore, with increasing the bottom wall temperature, the Euler number decreases which is due to absorption of more heat by the coolant and increasing bulk temperature of coolant and consequently reduction in viscosity of working fluid which reduce the pressure drop.

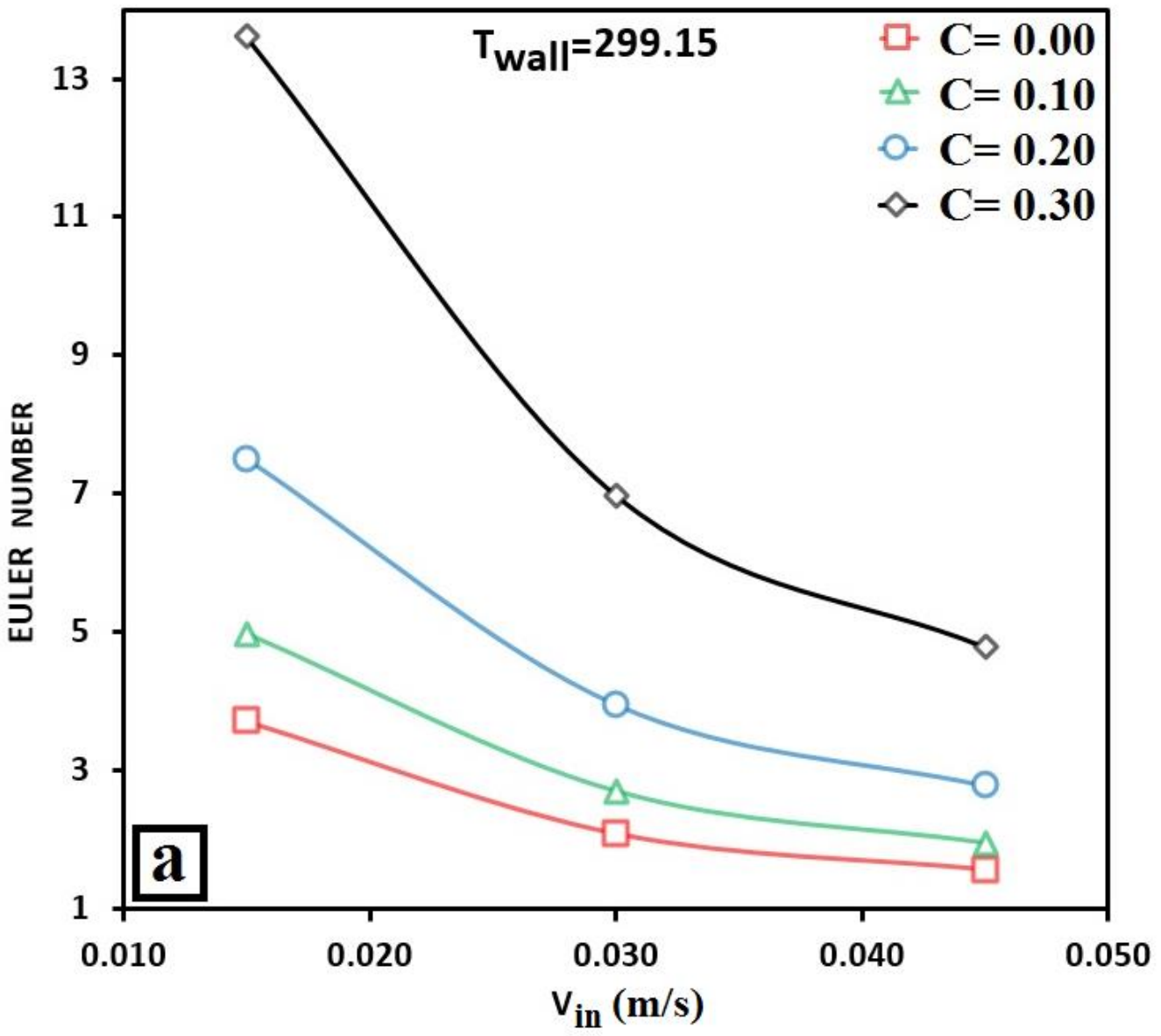

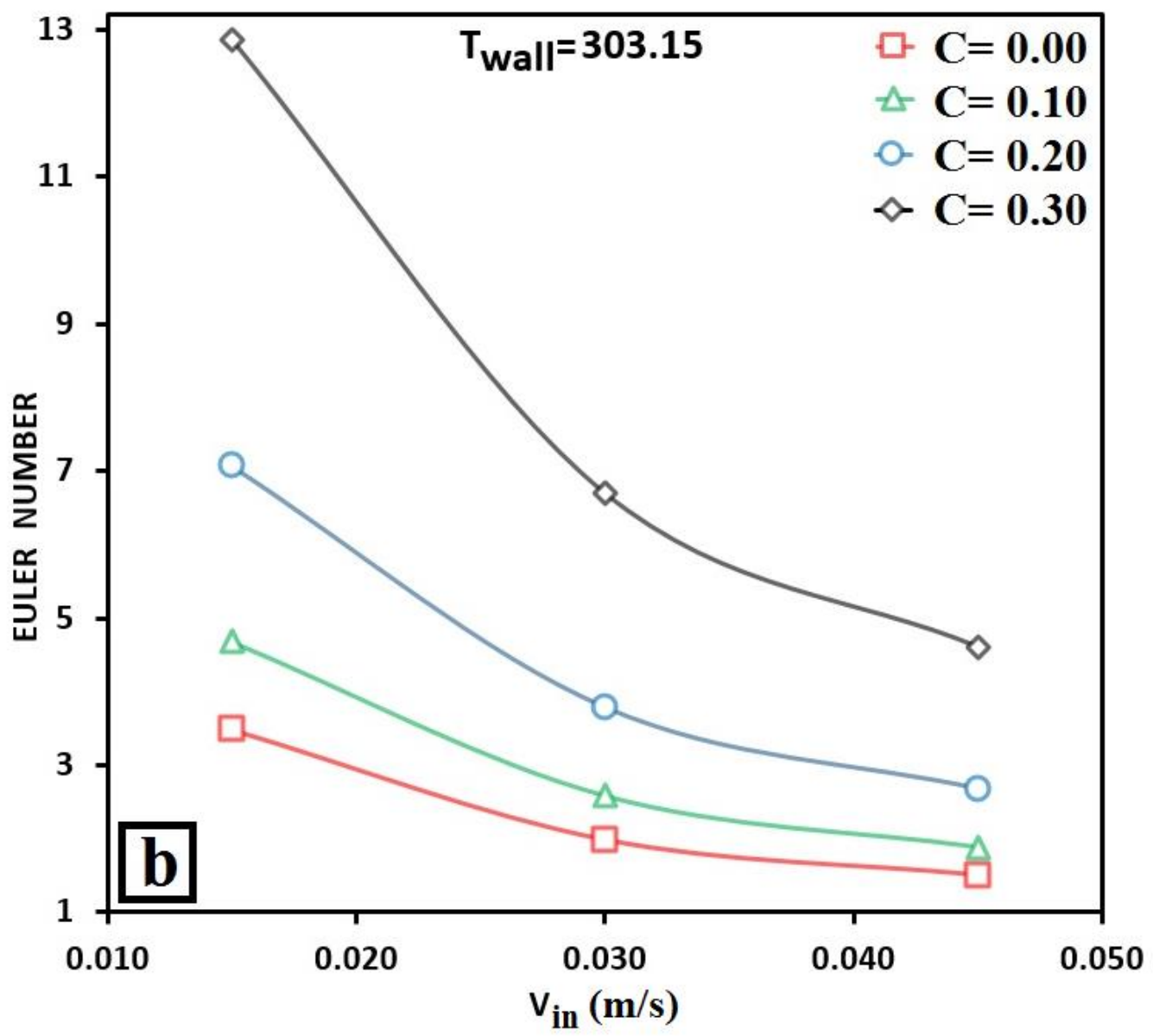

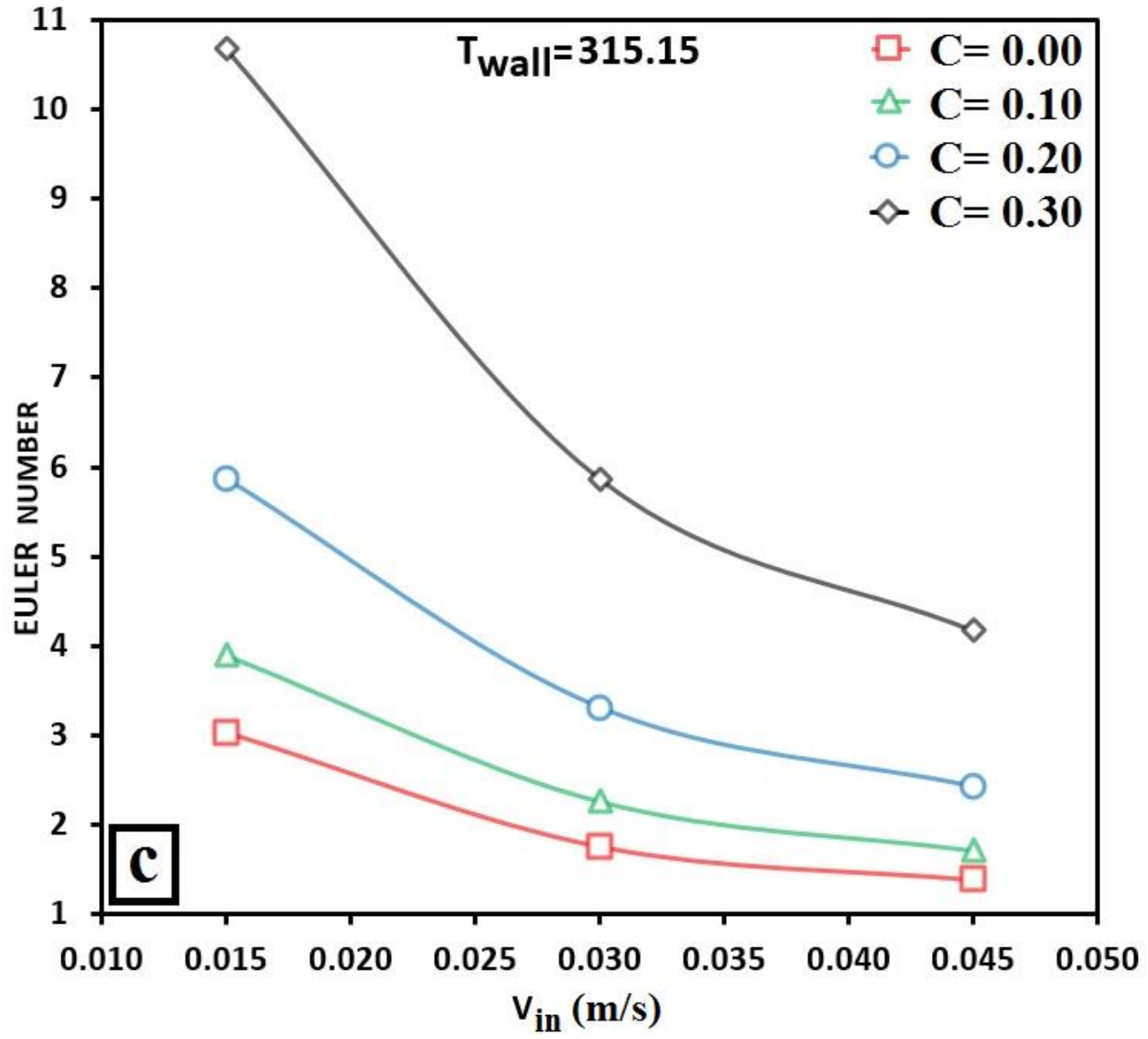

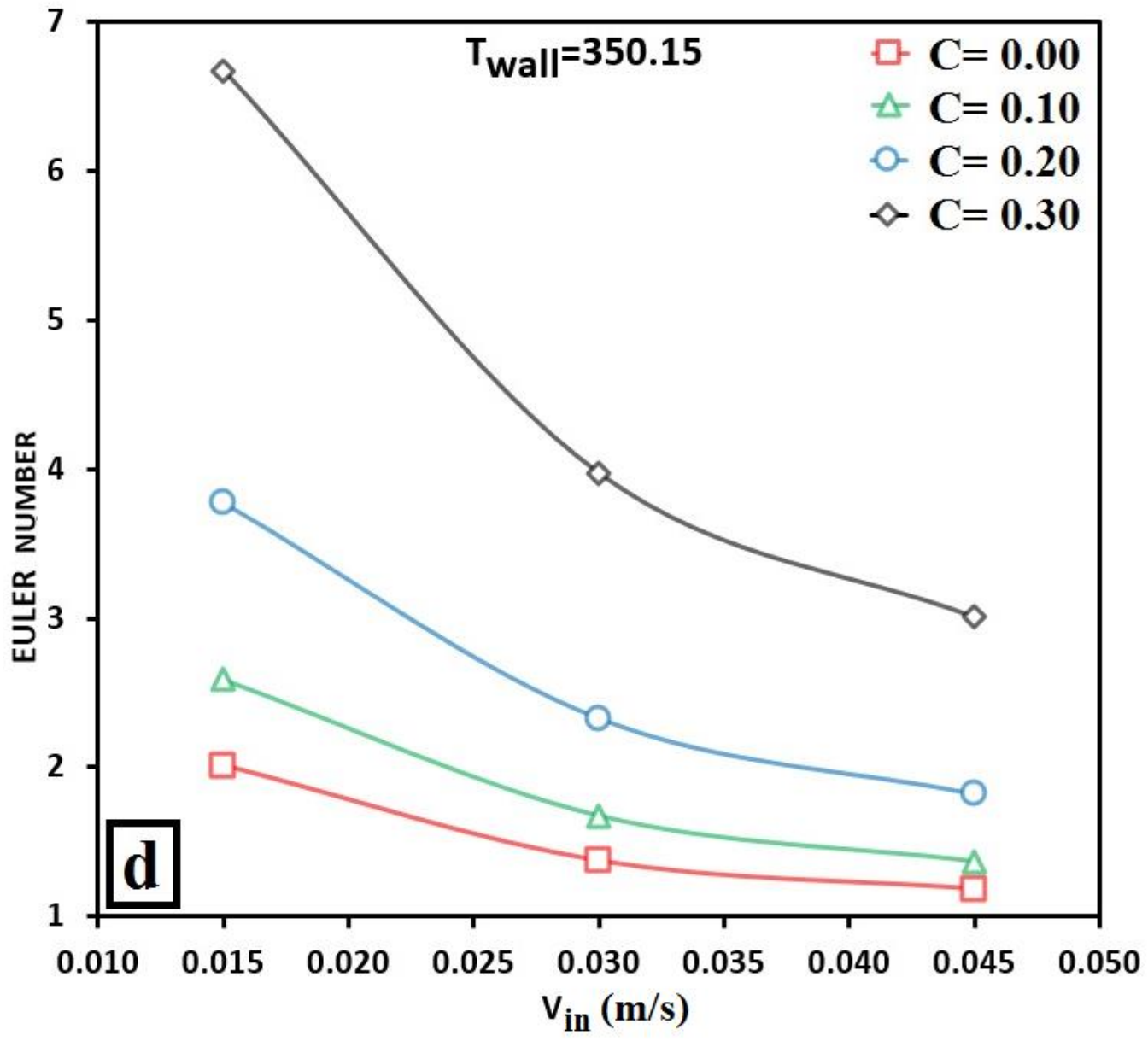

Figure 9. Effect of volume fraction on Euler number at various bottom wall temperatures.

Furthermore as seen in Figure 9, with using NEPCM slurry enhances the Euler number and consequently pressure drop, which is due to higher viscosity of slurry coolant compared to pure water. As seen, NEPCM slurry coolant is suffering from the increase in Euler number by increasing the NEPCM volume fraction because of higher viscosity compared with the pure water. It can be seen in viscosity contours shown in Figure 10 that the higher temperature of the coolant in regions close to the fin surfaces leads to the lower viscosity of the slurry. The viscosity of the slurry is calculated based on the below equation:

$$\mu_{eff} = \mu_w (1 - C - 1.16\, C^2)^{-2.5} \tag{9}$$

where, $\mu_w$ is the viscosity of water which is considered a temperature function [29]. The reason behind this phenomenon is that the molecules in colder segments are enjoying a lower energy level compared with hot segments. These low energetic molecules are more sluggish and cause increase in kinetic energy of each molecule. Therefore, the attractive /repulsive intermolecular forces become more dominant and result in a higher viscosity. The increase of the nanoparticles' volume fraction has a direct relation with viscosity. In other words, at higher volume fractions, the disturbance of fluid around the individual particles becomes stronger and it affects the fluid's internal shear stress. Accordingly, the higher internal shear stress of the fluid causes the liquid to be more sluggish and viscous.

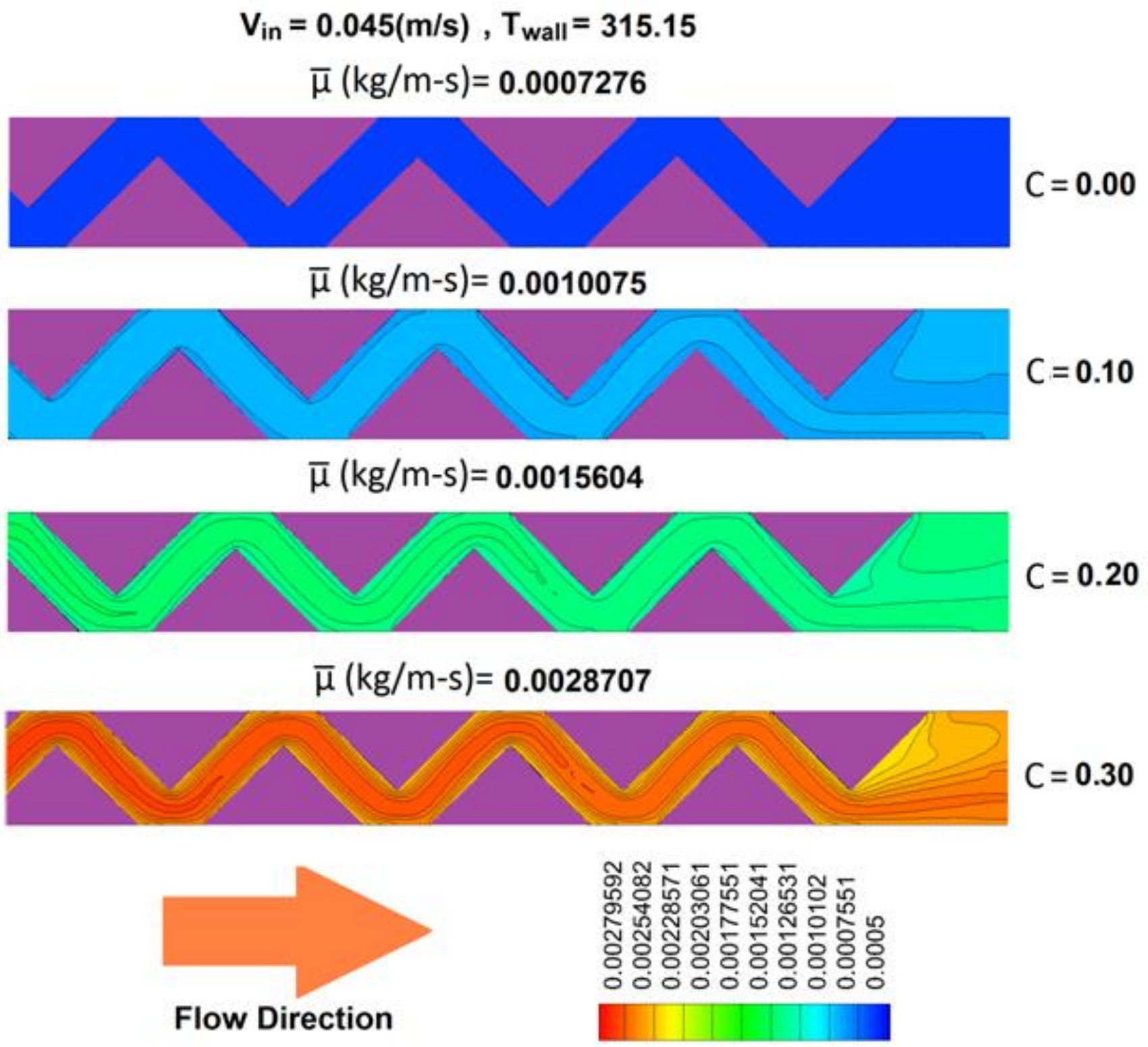

Figure 10. The distribution of viscosity of slurry in a plane cutting the fins at the middle of height computational domain for last seven fins, for a bottom wall temperature of 315.15 K and inlet velocity of 0.045 m/s at various volume fractions. (Unit: kg/m.s)

## Conclusion

A MPFHS-cooled NEPCM slurry was investigated to identify the effects of the nanoparticles volume fraction, inlet velocity, and wall temperature on the thermal and hydrodynamic performance in the laminar flow regime. It is demonstrated that the addition of NEPCM particles to the base fluid can enhance the Nusselt number remarkably but it has a drastic effect on Euler number; with increasing the volume fraction and decreasing Reynolds number, more increases on Euler number is observed. The results also show that increasing volume fraction and inlet velocity causes significant enhancement in Nusselt number but with increasing bottom wall temperature, the Nusselt number first increases and then decreases. Therefore, when the desirable 2.27, 1.81, 1.56 times higher maximum Nusselt numbers may be achieved when NEPCM slurries (C = 0.3) with $V_{in}$ = 0.015, 0.030, 0.045 (m/s) are employed, respectively, the more than 3 times greater associated Euler numbers denote the inevitable need for higher pumping power facilities, as well.

# Acknowledgement

Support for this work by the U.S. National Science Foundation under grant number CBET-1404482 is gratefully acknowledged.

**Nomenclature**

| | | |
|---|---|---|
| u | velocity in flow direction | (m $s^{-1}$) |
| h | heat transfer coefficient | (W $m^{-2}$ $K^{-1}$) |
| k | thermal conductivity | (W $m^{-1}$ $K^{-1}$) |
| $C_p$ | specific heat | (J/kg K) |
| H | height of pin fin | (m) |
| W | width | (m) |
| L | length of pin fin region | (m) |
| Nu | Nusselt number | |
| Re | Reynolds number | |
| Eu | Euler Number | |
| q | heat flux | (W $m^{-2}$) |
| V | velocity | (m $s^{-1}$) |
| Q | Volume flow rate | ($m^3$ $s^{-1}$) |
| $S_T$ | transverse distance of pin fins | (m) |
| $S_L$ | stream wise distance between pin fins | (m) |
| $\Delta p$ | pressure drop | (Pa) |
| T | temperature | (K) |
| C | volume fraction of nanoparticles | (%) |
| Greek symbols | | |
| $\rho$ | density | (kg $m^{-3}$) |
| $\mu$ | viscosity | (Pa s) |

Subscripts

| | |
|---|---|
| wall | Surface/wall |
| f | fluid |
| c | critical |
| m | average |

## References:

[1] Go, J. S., Kim, S. J., Lim, G., Yun, H., Lee, J., Song, I., and Pak, Y. E., 2001, "Heat transfer enhancement using flow-induced vibration of a microfin array," Sensors and Actuators A: Physical, 90(3), pp. 232-239.
[2] Peles, Y., Koşar, A., Mishra, C., Kuo, C.-J., and Schneider, B., 2005, "Forced convective heat transfer across a pin fin micro heat sink," International Journal of Heat and Mass Transfer, 48(17), pp. 3615-3627.
[3] Koşar, A., Mishra, C., and Peles, Y., 2005, "Laminar flow across a bank of low aspect ratio micro pin fins," Journal of Fluids Engineering, 127(3), pp. 419-430.
[4] Kosar, A., 2006, "Heat and Fluid Flow in MEMS-Based Pin Fin Heat Sinks," New York: Rensselaer Polytechnic Institute.
[5] Koşar, A., and Peles, Y., 2006, "Convective flow of refrigerant (R-123) across a bank of micro pin fins," International Journal of Heat and Mass Transfer, 49(17), pp. 3142-3155.
[6] Qu, W., and Siu-Ho, A., 2008, "Liquid Single-Phase Flow in an Array of Micro-Pin-Fins—Part II: Pressure Drop Characteristics," Journal of heat transfer, 130(12), p. 124501.
[7] Seyf, H. R., and Feizbakhshi, M., 2012, "Computational analysis of nanofluid effects on convective heat transfer enhancement of micro-pin-fin heat sinks," International Journal of Thermal Sciences, 58, pp. 168-179.
[8] Liu, M., Liu, D., Xu, S., and Chen, Y., 2011, "Experimental study on liquid flow and heat transfer in micro square pin fin heat sink," International Journal of Heat and Mass Transfer, 54(25), pp. 5602-5611.
[9] Sabbah, R., Seyed-Yagoobi, J., and Al-Hallaj, S., 2011, "Heat Transfer Characteristics of Liquid Flow With Micro-Encapsulated Phase Change Material: Numerical Study," Journal of Heat Transfer, 133(12), pp. 121702-121702.
[10] Hao, Y.-L., and Tao, Y., 2004, "A numerical model for phase-change suspension flow in microchannels," Numerical Heat Transfer, Part A: Applications, 46(1), pp. 55-77.
[11] Wang, X., Niu, J., Li, Y., Zhang, Y., Wang, X., Chen, B., Zeng, R., and Song, Q., 2008, "Heat transfer of microencapsulated PCM slurry flow in a circular tube," AIChE journal, 54(4), pp. 1110-1120.
[12] Kuravi, S., Kota, K. M., Du, J., and Chow, L. C., 2009, "Numerical Investigation of Flow and Heat Transfer Performance of Nano-Encapsulated Phase Change Material Slurry in Microchannels," Journal of Heat Transfer, 131(6), pp. 062901-062901.
[13] Sabbah, R., Seyed-Yagoobi, J., and Al-Hallaj, S., 2012, "Natural convection with micro-encapsulated phase change material," Journal of Heat Transfer, 134(8), p. 082503.
[14] Kondle, S., Alvarado, J. L., and Marsh, C., 2013, "Laminar Flow Forced Convection Heat Transfer Behavior of a Phase Change Material Fluid in Microchannels," Journal of Heat Transfer, 135(5), pp. 052801-052801.
[15] Seyf, H. R., Zhou, Z., Ma, H., and Zhang, Y., 2013, "Three dimensional numerical study of heat-transfer enhancement by nano-encapsulated phase change material slurry in microtube heat sinks with tangential impingement," International Journal of Heat and Mass Transfer, 56(1), pp. 561-573.
[16] Seyf, H. R., Wilson, M. R., Zhang, Y., and Ma, H., 2014, "Flow and Heat Transfer of Nanoencapsulated Phase Change Material Slurry Past a Unconfined Square Cylinder," Journal of Heat Transfer, 136(5), p. 051902.
[17] Yang, J., Zeng, M., Wang, Q., and Nakayama, A., 2010, "Forced Convection Heat Transfer Enhancement by Porous Pin Fins in Rectangular Channels," Journal of Heat Transfer, 132(5), pp. 051702-051702.
[18] Rajabi Far, B., Mohammadian, S. K., Khanna, S. K., and Zhang, Y., 2015, "Effects of pin tip-clearance on the performance of an enhanced microchannel heat sink with oblique fins and phase change material slurry," International Journal of Heat and Mass Transfer, 83, pp. 136-145.
[19] Zhang, Y., and Faghri, A., 1995, "Analysis of forced convection heat transfer in microencapsulated phase change material suspensions," Journal of Thermophysics and Heat Transfer, 9(4), pp. 727-732.

[20] Alisetti, E. L., and Roy, S. K., 2000, "Forced Convection Heat Transfer to Phase Change Material Slurries in Circular Ducts," Journal of Thermophysics and Heat Transfer, 14(1), pp. 115-118.
[21] Karnis, A., Goldsmith, H. L., and Mason, S. G., 1966, "The kinetics of flowing dispersions: I. Concentrated suspensions of rigid particles," Journal of Colloid and Interface Science, 22(6), pp. 531-553.
[22] Watkins, R. W., Robertson, C. R., and Acrivos, A., 1976, "Entrance region heat transfer in flowing suspensions," International Journal of Heat and Mass Transfer, 19(6), pp. 693-695.
[23] Rajabifar, B., 2015, "Enhancement of the performance of a double layered microchannel heatsink using PCM slurry and nanofluid coolants," International Journal of Heat and Mass Transfer, 88, pp. 627-635.
[24] Seyf, H. R., and Layeghi, M., 2010, "Numerical analysis of convective heat transfer from an elliptic pin fin heat sink with and without metal foam insert," Journal of Heat Transfer, 132(7), p. 071401.
[25] Van Doormaal, J., and Raithby, G., 1984, "Enhancements of the SIMPLE method for predicting incompressible fluid flows," Numerical heat transfer, 7(2), pp. 147-163.
[26] Leonard, B. P., 1995, "Order of accuracy of QUICK and related convection-diffusion schemes," Applied Mathematical Modelling, 19(11), pp. 640-653.
[27] Rhie, C., and Chow, W., 1983, "Numerical study of the turbulent flow past an airfoil with trailing edge separation," AIAA journal, 21(11), pp. 1525-1532.
[28] Goel, M., Roy, S. K., and Sengupta, S., 1994, "Laminar forced convection heat transfer in microcapsulated phase change material suspensions," International Journal of Heat and Mass Transfer, 37(4), pp. 593-604.
[29] Vand, V., 1948, "Viscosity of Solutions and Suspensions. I. Theory," The Journal of Physical and Colloid Chemistry, 52(2), pp. 277-299.